\documentclass[twocolumn]{aastex6}
\usepackage{graphicx}
\usepackage[tbtags]{amsmath}
\usepackage{dsfont}
\usepackage{amssymb,amsfonts,amstext,amsgen,amsopn,amsxtra,indentfirst,times,chemformula}
\usepackage{caption,subcaption}
\usepackage{csquotes}
\MakeOuterQuote{"}
\usepackage{xspace}
\usepackage[capitalise,noabbrev]{cleveref}
\usepackage{extdash}
\usepackage{savesym}
\savesymbol{tablenum}
\usepackage[separate-uncertainty]{siunitx}
\restoresymbol{SIX}{tablenum}
\usepackage{bm,upgreek,physics, interval}
\intervalconfig{soft open fences}

\usepackage{tikz}
\usetikzlibrary{external}
\newcommand{\prob}{\opbraces{\operatorname{p}}}

\newcommand{\given}{\,\vert\,}
\newcommand{\evidence}[1][m]{\opbraces{Z\ifblank{#1}{}{_{#1}}}}
\newcommand{\Expectation}[1][]{\opbraces{\operatorname{\mathds{E}\ifblank{#1}{}{_{#1}}}}}
\newcommand{\Normal}{\opbraces{\operatorname{Normal}}}
\newcommand{\nmod}{\ensuremath{N_{\textrm{mod}}}\xspace}

\newcommand{\aflux}[1][i]{\ensuremath{F\ifblank{#1}{}{_{#1}}}\xspace}
\newcommand{\flux}{\ensuremath{\bm{\mathrm{\aflux[]}}}\xspace}
\newcommand{\aerrtot}[1][i]{\ensuremath{\sigma\ifblank{#1}{}{_{#1}}}\xspace}
\newcommand{\errtot}{\ensuremath{\bm{\mathrm{\aerrtot[]}}}\xspace}
\newcommand{\aerrrep}[1][i]{\ensuremath{s\ifblank{#1}{}{_{#1}}}\xspace}
\newcommand{\errrep}{\ensuremath{\bm{\mathrm{\aerrrep[]}}}\xspace}
\newcommand{\aerradd}[1][]{\ensuremath{\varsigma\ifblank{#1}{}{_{#1}}}\xspace}

\newcommand{\adata}[1][i]{\ensuremath{D\ifblank{#1}{}{_{#1}}}\xspace}
\newcommand{\data}{\ensuremath{\bm{\mathrm{\adata[]}}}\xspace}
\newcommand{\tdata}{\ensuremath{\data_0}\xspace}

\newcommand{\nparams}{\ensuremath{\bm{\upnu}}\xspace}
\newcommand{\iparams}{\ensuremath{\bm{\uptheta}}\xspace}
\newcommand{\iparam}[1][]{\ensuremath{\theta\ifblank{#1}{}{_{#1}}}\xspace}
\newcommand{\Model}{\ensuremath{\mathcal{M}}\xspace}
\newcommand{\aModel}[1][m]{\ensuremath{M\ifblank{#1}{}{_{#1}}}\xspace}
\newcommand{\BF}[1][]{\opbraces{\operatorname{\mathrm{BF}}\ifblank{#1}{}{_{#1}}}}
\newcommand{\nreratio}{\opbraces{\operatorname{r}}}
\newcommand{\nreapprox}[1][]{\opbraces{\operatorname{\hat{r}}\ifblank{#1}{}{_{#1}}}}

\newcommand{\acronym}{\texttt{FASTER}}

\begin{document}

\renewcommand{\edit}[1]{\textcolor{red}{#1}}

\title{Near-instantaneous Atmospheric Retrievals and Model Comparison with \acronym}

\author{Anna Lueber\altaffilmark{1,2}}
\author{Konstantin Karchev\altaffilmark{3}}
\author{Chloe Fisher\altaffilmark{4}}
\author{Matthias Heim\altaffilmark{1}}
\author{Roberto Trotta\altaffilmark{3,5,6,7}}
\author{Kevin Heng\altaffilmark{1,8,9,10}}
\altaffiltext{1}{Faculty of Physics, Ludwig Maximilian University, Scheinerstrasse 1, D-81679, Munich, Bavaria, Germany}
\altaffiltext{2}{Center for Space and Habitability, University of Bern, Gesellschaftsstrasse 6, CH-3012 Bern, Switzerland}
\altaffiltext{3}{Theoretical and Scientific Data Science, Scuola Internazionale Superiore di Studi Avanzati (SISSA), via Bonomea 265, 34136 Trieste, Italy}
\altaffiltext{4}{Department of Physics, University of Oxford, Keble Road, Oxford, OX1 3RH, United Kingdom}
\altaffiltext{5}{Astrophysics Group, Physics Department, Blackett Lab, Imperial College London, Prince Consort Road, London SW7 2AZ, United Kingdom}
\altaffiltext{6}{INFN – National Institute for Nuclear Physics, Via Valerio 2, 34127 Trieste, Italy}
\altaffiltext{7}{Italian Research Center on High-Performance Computing, Big Data and Quantum Computing}
\altaffiltext{8}{ARTORG Center for Biomedical Engineering Research, University of Bern, Murtenstrasse 50, CH-3008, Bern, Switzerland}
\altaffiltext{9}{University College London, Department of Physics \& Astronomy, Gower St, London, WC1E 6BT, United Kingdom}
\altaffiltext{10}{Astronomy \& Astrophysics Group, Department of Physics, University of Warwick, Coventry CV4 7AL, United Kingdom}

\begin{abstract}
In the era of the James Webb Space Telescope (JWST), the dramatic improvement in the spectra of exoplanetary atmospheres demands a corresponding leap forward in our ability to analyze them: atmospheric retrievals need to be performed on thousands of spectra, applying to each large ensembles of models (that explore atmospheric chemistry, thermal profiles and cloud models) to identify the best one(s). In this limit, traditional Bayesian inference methods such as nested sampling become prohibitively expensive.  We introduce \texttt{FASTER} (Fast Amortized Simulation-based Transiting Exoplanet Retrieval), a neural-network based method for performing atmospheric retrieval and Bayesian model comparison at a fraction of the computational cost of classical techniques.  We demonstrate that the marginal posterior distributions of all parameters within a model as well as the posterior probabilities of the models we consider match those computed using nested sampling both on mock spectra, and for the real NIRSpec PRISM spectrum of WASP-39b.  The true power of the \texttt{FASTER} framework comes from its amortized nature, which allows the trained networks to perform practically instantaneous Bayesian inference and model comparison over ensembles of spectra -- real or simulated -- at minimal additional computational cost.  This offers valuable insight into the expected results of model comparison (e.g., distinguishing cloudy from cloud-free and isothermal from non-isothermal models), as well as their dependence on the underlying parameters, which is computationally unfeasible with nested sampling.  This approach will constitute as large a leap in spectral analysis as the original retrieval methods based on Markov Chain Monte Carlo have proven to be.
\end{abstract}


\section{Introduction}
\label{sect:intro}

Measuring spectra of the atmospheres of exoplanets has become routine (e.g., \citealt{carter24,fu25,kirk25}).  Encoded within these spectra is information on the physical and chemical properties of an atmosphere, which may be retrieved using Bayesian inference.  Atmospheric retrieval is the approach of solving this inverse problem to extract the posterior distributions of atomic/molecular abundances, thermal profiles, cloud/haze\footnote{For the current study, we use the terms ``cloud'' and ``haze'' synonymously.} properties and other features of an atmosphere from its measured spectrum \citep{bh20}.  It is a crucial tool for utilizing atmospheres as chemical probes of exoplanets via remote sensing.

Since its introduction to the exoplanet literature by \cite{ms09}, atmospheric retrieval has been applied to transiting exoplanets \citep{bs12,line13,waldmann15}, directly imaged exoplanets \citep{lee13} and brown dwarfs \citep{line15}.  Among likelihood-based Bayesian methods, nested sampling is one of the most widely employed techniques to obtain posterior samples (e.g. \citealt{bs13}), while in more recent years machine learning methods have been used to perform atmospheric retrieval as well (e.g., \citealt{mn18,zw18,cobb19,yip21,ardevolmartinez22,yip24}). Currently used approaches face two major challenges: firstly, several models may be compatible with a given observed spectrum, and it would be useful to be able to compare several models at once with the data, in order to select the ``best'' (as measured by the Bayesian evidence), or else perform model averaging \citep[e.g.,][]{MacKay2003} to obtain more robust constraints. Secondly, while it is straightforward to write a Bayesian hierarchical model \citep[e.g.,][]{Gelman2014} to study the parameters describing the population distribution of exoplanetary atmospheres, current Bayesian posterior samplers do not scale to the high-dimensional parameter space that such an approach would require. 

Speaking to the first challenge, cogent motivation to move beyond traditional atmospheric retrieval approaches came from a recent study by \cite{lueber24}, who analyzed four JWST spectra of the benchmark hot Jupiter WASP-39b.  For each of these spectra, a family of models considered various combinations of 7 chemical species, 4 treatments of temperature-pressure profiles and both gray\footnote{Gray clouds have constant cross sections and consist of particles with sizes exceeding the wavelength of radiation being probed.} and non-gray\footnote{Non-gray clouds consist of particles with sizes smaller than the wavelength of radiation being probed and have wavelength-dependent cross sections.} clouds.  This amounted to about 400 retrievals for analyzing these spectra of WASP-39b alone, even without considering vertically non-uniform chemical abundance profiles or more sophisticated cloud models. This amounted to a total computational budget of $\sim 10^4$
GPU-hours. Bayesian model comparison among all 400 models was performed to identify the simplest model giving a good explanation of the data (a quantitative implementation of Occam's razor) -- clearly something that cannot be carried out manually for a population of hundreds of exoplanets.

In the current study, we introduce a new approach to atmospheric retrieval: simulation-based inference \citep[SBI; for methodological reviews, see][]{Cranmer_2020,Lueckmann_2021}. While SBI has recently gained popularity across various scientific domains\footnote{A list of applications is automatically being compiled at \url{https://simulation-based-inference.org/}.}, it has seen only limited application in the analysis of exoplanetary atmospheres \citep{vasist23,aubin23,ardevolmartinez24,gebhard25}. We demonstrate that SBI can:

\begin{itemize}
\item Learn a probabilistic model from a large training suite of forward simulations of spectra that include physics of varying complexity (different cloud parametrizations, different treatments of temperature-pressure profiles, etc) for any parameter combination within a prior box;

\item Once trained, perform almost-instantaneous Bayesian inference on mock spectra and deliver posterior distributions of parameters that are consistent with a likelihood-based Bayesian analysis using nested sampling;

\item Perform near-instantaneous Bayesian model comparison over the entire ensemble of models being considered with minimal upfront computational cost;

\item Quantify the average performance of Bayesian model selection over the entire prior range considered, an analysis that is computationally impossible using traditional retrieval methods as it requires thousands of retrievals per model;

\item Reproduce the WASP-39b retrieval analysis of \cite{lueber24}, which was performed using nested sampling.

\end{itemize}

In Section \ref{sect:methods}, we describe our methodology, including our forward model (for computing synthetic spectra) and likelihood-based versus SBI approaches.  In Section \ref{sect:results}, we present our results.  In Section \ref{sect:discussion}, we discuss the implication of our findings, compare to previous work and suggest opportunities for future work.

\section{Methodology}
\label{sect:methods}

\subsection{Forward model\label{sect:forward}}

A central ingredient in any atmospheric retrieval framework is the forward model that takes input parameters and computes a spectrum. A transmission spectrum is computed by determining the wavelength-dependent transit chord that has an optical depth of about unity \citep{fortney05}, which is performed using ray-tracing through the atmosphere \citep{brown01}.

We assume a one-dimensional, plane-parallel atmosphere with 99 discrete layers spanning pressures from $P=10$ bar to 1 $\mu$bar equally spaced in $\log{P}$.  The temperature-pressure profile is described using a finite element approach, which allows smooth, continuous profiles to be constructed using a small number of parameters (see Section 2.5 of \citealt{kitzmann20}).  In the current study, the temperature-pressure profile uses between one and four parameters, where a single-parameter description corresponds to an isothermal transit chord.  Two, three and four parameters allow for one, two and three linear slopes in the temperature-pressure profile, respectively.  This forward model and temperature-pressure profile parametrization have previously been implemented in the open-source \texttt{BeAR} retrieval code\footnote{\url{https://newstrangeworlds.github.io/BeAR}} (originally named \texttt{HELIOS-R2}; \citealt{kitzmann20}), which we use for the current study. 

Indispensable inputs are the cross sections of atoms and molecules as functions of wavelength, temperature and pressure.  These cross sections are computed using spectroscopic line lists as inputs \citep{gh15,grimm21} and at a spectral resolution of 0.01 cm$^{-1}$, but are downsampled to 1 cm$^{-1}$ when performing atmospheric retrievals.  The line lists for water (H$_2$O; \citealt{polyansky2018}), carbon dioxide (CO$_2$; \citealt{tashkun2011}), carbon monoxide (CO; \citealt{li2015}), sulfur dioxide (SO$_2$; \citealt{underwood2016}) and hydrogen sulfide (H$_2$S; \citealt{azzam2016}) are used.  Cross sections for sodium (Na) and potassium (K) are taken from \cite{kitzmann20}.  Collision-induced absorption (CIA) associated with hydrogen-hydrogen (H$_2$-H$_2$; \citealt{abel2011}) and hydrogen-helium (H$_2$-He; \citealt{abel2012}) pairs is also included.  These cross sections are publicly available through the DACE database\footnote{\url{https://dace.unige.ch}} \citep{grimm21}.  Each cross section is multiplied by the volume mixing ratio (relative abundance by number) of the respective atom or molecule when computing the total absorption cross section.

Atmospheres are generally expected to contain condensates or aerosols, often termed ``clouds'' or ``hazes''.  While their composition is connected to that of the gas in the atmosphere, this chemical relationship is typically not modeled in retrievals.  Rather, the cross section of the cloud is parametrized.  For gray clouds, we parametrize the cross section using a constant optical depth.  For non-gray clouds consisting of spherical particles with a radius $r_{\rm cloud}$, the cross section is given by $Q \pi r_{\rm cloud}^2$ where the extinction efficiency is given by equation (32) of \cite{kh18},
\begin{equation}
Q \propto \frac{\tau_{\rm cloud}}{Q_0 x^{-a_0} + x^{0.2}}.
\end{equation}
The cloud optical depth (referenced to its value at 1 $\mu$m) is given by $\tau_{\rm cloud}$.  The size parameter is $x = 2 \pi r_{\rm cloud}/\lambda$, where $\lambda$ is the wavelength of radiation being probed.  The slope index $a_0$ describes the slope of $Q(x)$ in the small-particle regime.  The parameter $Q_0$ is a proxy for the particle composition \citep{kh18}, but is typically unconstrained in retrievals (e.g., \citealt{lueber24}).  In both gray and non-gray cases, we assume the cloud to be vertically semi-infinite in extent, which is appropriate for transmission spectra (as they do not probe that deeply into the atmosphere) and parameterize the top boundary using a ``cloud-top pressure'' ($P_{\rm cloud}$).

We consider $\nmod=12$ distinct simulator configurations (\emph{models}), formed by picking one among 4 temperature--pressure parametrizations ("TP1", "TP2", "TP3", "TP4") and one among 3 cloud models (cloudfree: "CF", gray clouds: "G", and non-gray clouds: "NG"). Each model has a different  number of parameters, ranging from 12 for the simplest model (TP1 CF: $\qty{T_0, \log{g}, r_p, r_s, X_{\ch{H2O}}, X_{\ch{CO}}, X_{\ch{CO2}}, X_{\ch{H2S}}, X_{\ch{K}}, X_{\ch{Na}}, X_{\ch{SO2}}, \aerradd}$) to 20 for the most complex model (TP4 NG). Each model features the TP1 CF set of 12 parameters, plus additional parameters that describe more complex temperature-pressure profiles (up to 3 additional parameters) and cloud parameters (up to 5 additional parameters; see Table \ref{tab:NS_priors}). In all cases, we synthesize a (noiseless) transmission spectrum \flux (in parts-per-million: \si{ppm}) in the range \SIrange[range-phrase=--, range-units=single]{0.53}{5.34}{\micro\meter} with \num{207} wavelength bins, as appropriate for the PRISM mode of JWST's NIRSpec instrument.

\subsection{Statistical model}

 The observable \data is a noisy measurement of the noiseless spectrum  \flux with Gaussian uncertainties \errtot, which we assume independent across wavelength bins:
\begin{equation}\label{eqn:like}
    \adata \sim \Normal(\aflux, \aerrtot^2).
\end{equation}
While the data-reduction procedure reports an error \emph{estimate} \errrep, we allow for an additional contribution\footnote{Previously called "error inflation" in \citet{kitzmann20} (and written as $10^\epsilon \equiv \aerrtot[]^2$), this additional variance is an ad-hoc component necessitated by the imperfections of even the best likelihood-based fits. While in principle, it can be increased in complexity by considering \emph{different} additional noise levels \aerrtot across the spectrum or having them be correlated (e.g., as would arise from an "incorrect" subtraction of the star spectrum or improper calibration), the proper way to address is by improving the explicit model both on the side of the physics and chemistry of the atmosphere and of the instrument.} \aerradd to account both for potential mismodelling of instrumental effects and for the fact that currently, no forward model can account for all the physics and chemistry present in a real atmosphere \citep{kitzmann20}. The two components are added in quadrature to give the total variance for the likelihood of \cref{eqn:like}:
\begin{equation} \label{eqn:noise}
    \aerrtot^2 = \aerrrep^2 + \aerradd^2.
\end{equation}
Since the reported uncertainties in JWST observations are on the order of \SI{100}{ppm}, we expect a similar scale for \aerradd. Nevertheless, we treat it as a free model parameter and infer it together with the other quantities in the simulator, after placing a wide prior that encompasses both the case that \errrep fully captures the uncertainties, as well as the possibility that they are grossly underestimated.

This and the remaining priors we use are listed in \cref{tab:NS_priors}. For the most part, they are broad (log-)uniform distributions, except for the planetary radius, surface gravity, and the stellar radius, for which we adopt Gaussian prior constraints from external measurements of the hot Jupiter WASP-39b \citep{mancini18}.  This allows for comparison with the previous analysis by \citet{lueber24}, which we reproduce for this study using the traditional nested sampling technique \citep{skilling06} implemented as \texttt{MultiNest}\footnote{Everywhere, \texttt{MultiNest} is run with $\mathtt{nlive} = \num{1000}$ live points and tolerance of $\Delta \ln Z = \num{0.5}$.} \citep{feroz09} within the open-source \texttt{BeAR} code. We consider the resulting parameter posteriors and model evidences as the ground truth with which to validate our methodology.

\begin{table*}[ht]
\centering
\caption{Simulator parameters, their prior distributions, and mock values$^\spadesuit$ used.\label{tab:NS_priors}}
\begin{tabular}{lccccc}
\hline
Parameter & Symbol & Mock value$^\spadesuit$ & Prior range & Distribution & Units \\ \hline

Planetary surface gravity$^\ddagger$ & $\log{g}$
    & \num{2.67} & \num{2.629\pm0.051}
    & Gaussian & \si{\centi\meter\per\second\squared} \\
Planetary radius$^\ddagger$ & $r_p$
    & \num{1.28} & \num{1.279\pm0.051}
    & Gaussian & $R_{\rm J}$ \\
Stellar radius$^\ddagger$ & $r_s$
    & \num{0.95} & \num{0.939\pm0.030}
    & Gaussian & $R_\odot$ \\

Volume mixing ratios\textsuperscript{*} & $X_{i}$
    & --- & $\interval{\num{e-12}}{\num{e-1}}$
    & Log-uniform & --- \\

Temperature & $T_0$
    & \num{1000} & $\interval{\num{500}}{\num{3000}}$
    & Uniform & \si{\kelvin} \\
Slope between adjacent temperature nodes$^\dagger$ & $b_{i=1\dots4}$
    & --- & $\interval{\num{0.1}}{\num{3.0}}$
    & Uniform & --- \\
Additional noise$^\clubsuit$ & $\varsigma$
    & \num{0} & $\interval{\num{4}}{\num{420}}$
    & Log-uniform & \si{ppm} \\

\hline
\textit{Gray clouds} \\
Cloud-top pressure & $P_{\rm cloud}$
    & \num{e-3} & $\interval{\num{e-6}}{\num{e1}}$ & Log-uniform & \si{\bar} \\
Optical depth & $\tau_{\rm cloud}$
    & 500 & $\interval{\num{e-5}}{\num{e3}}$ & Log-uniform & --- \\

\hline
\textit{Non-gray clouds} \\
Cloud-top pressure & $P_{\rm cloud}$
    & --- & $\interval{\num{e-6}}{\num{e1}}$
    & Log-uniform & \si{\bar} \\
Reference optical depth & $\tau_{\rm cloud}$ 
    & --- & $\interval{\num{e-5}}{\num{e3}}$ & Log-uniform & --- \\
Composition parameter & $Q_0$ 
    & --- & $\interval{\num{1}}{\num{100}}$ & Uniform & --- \\
Slope index & $a_0$ 
    & --- & $\interval{\num{3}}{\num{6}}$ & Uniform & --- \\
Spherical cloud particle radius & $r_{\rm cloud}$ 
    & --- & $\interval{\num{e-7}}{\num{e-1}}$ & Log-uniform & \si{\centi\meter} \\

\hline
\end{tabular}\\
\scriptsize{
\textsuperscript{*} Values used for the mock retrievals are $X_{\ch{H2O}} = \num{e-3}$, $X_{\ch{CO}} = \num{e-3}$, $X_{\ch{CO2}} = \num{e-4}$, $X_{\ch{H2S}} = \num{e-4}$, $X_{\ch{K}} = \num{e-7}$, $X_{\ch{Na}} = \num{e-4}$ and $X_{\ch{SO2}} = \num{e-6}$.\\
$\spadesuit$: Parameter values assumed for the mock retrievals.\\
$\dagger$: For the non-isothermal profiles, $b_i$ is the slope between two adjacent temperature nodes \citep{kitzmann20}.\\
$\ddagger$: Based on measured values reported in \cite{mancini18}.\\
$\clubsuit$: We consider additional variance ranging from $\aerrrep[\rm min]^2/100$ to $4\aerrrep[\rm max]^2$, where \aerrrep[\rm min] and \aerrrep[\rm max] are the minimum \\ and maximum values, respectively, of the reported observational uncertainties \aerrrep.}
\end{table*}

\subsection{Simulation-based inference}

Simulation-based inference\footnote{For reviews and a comprehensive comparison of methods, see \cite{Cranmer_2020} and \cite{Lueckmann_2021}. Extensive up-to-date lists of references to software and applications are kept at \url{https://github.com/smsharma/awesome-neural-sbi} and \url{https://simulation-based-inference.org}.} is an emerging comprehensive framework for Bayesian analysis that has been rapidly gaining popularity in recent years, alongside developments in machine learning and deep neural networks (NNs). SBI (also called "likelihood-free inference") uses a stochastic model as a forward simulator that samples parameter values from given priors and produces plausible mock data, which are then used to train a NN to perform inference without the need to explicitly evaluate the likelihood function. The main advantages of SBI over likelihood-based Bayesian techniques are its ability to learn a complicated (possibly intractable) likelihood directly from mock data; its scalability to a large number of free parameters (implicitly marginalized in the simulator); and -- crucial for this work -- its ability to perform almost instantaneous parameter inference and model comparison once appropriately trained.

\subsubsection{Marginal parameter inference with neural ratio estimation
\label{sec:nre}}

Any likelihood-based approach requires \emph{joint} inference over the full set of model parameters, which can be computationally demanding even for parameter spaces of moderate dimension. However, in many cases, one is only interested in the \emph{marginal} posterior of a few parameters of interest \iparams, with the remaining (nuisance) parameters \nparams integrated out from the joint posterior:
\begin{equation}\begin{split}\label{eqn:post}
    \prob(\iparams\given\tdata)
    & \propto \prob(\iparams) \times \prob(\tdata\given\iparams)
    \\ & \propto \prob(\iparams) \times \int \prob(\tdata \given \nparams, \iparams) \prob(\nparams\given\iparams)  \dd{\nparams},
\end{split}\end{equation}
where \tdata is the observed data. In SBI, this integration is performed implicitly by stochastically sampling \nparams while simulating training examples. While there exist several conceptually distinct variants of SBI \citep[see e.g.,][]{Papamakarios_2016, Papamakarios_2019,Lueckmann_2019}, here we focus on neural ratio estimation \citep[NRE;][]{Hermans_2020}: in this approach, the problem of inferring the posterior distribution is converted into the simpler task of binary classification, which neural networks are particularly adept for. Given parameters of interest, \iparams, and simulated data, \data, the network is trained to approximate the ratio
\begin{equation}\label{eqn:nreratio}
    \nreratio(\iparams, \data) \equiv \frac{\prob(\iparams, \data)}{\prob(\iparams) \prob(\data)} = \frac{\prob(\data\given\iparams)}{\prob(\data)} = \frac{\prob(\iparams\given\data)}{\prob(\iparams)},
\end{equation}
where the last equality follows from Bayes's theorem and shows that one can obtain the posterior in \cref{eqn:post} by simply multiplying the prior density $\prob(\iparams)$ by $\nreratio(\iparams, \tdata)$ evaluated at the observed data. The key realization in NRE is that a neural network can be trained to approximate equation (\ref{eqn:nreratio}) through binary classification\footnote{Specifically, the network outputs a single real number $\ln\nreapprox(\iparams, \data)$ and is trained using the standard binary cross-entropy (BCE) loss functional.} of pairs $\qty(\iparams, \data)$ into two classes: one where \iparams are the parameter from which \data was generated ("joint" pairs from $\prob(\iparams, \data)$), and another where \iparams are randomly drawn from the prior ("marginal" pairs from $\prob(\iparams)\prob(\data)$).

In general, $\iparams$ can be an arbitrary group of parameters, although in the following we will specialize to a single parameter of interest, \iparam. This delivers the set of marginal one-dimensional posteriors, which are usually sufficient for drawing scientific conclusions. It is also easy to obtain a full "triangle plot", showing two-dimensional joint marginal posteriors among pairs of parameters, by having \iparams represent each pair of parameters of interest \citep[see e.g.][Figure 8]{Cole_2022}. Moreover, the so-called auto-regressive extension of NRE \citep{AnauMontel_2024} can derive a full joint posterior over many ($\gtrsim 20$) parameters of interest with the same simulation and training-time budget. These techniques will be employed in future in-depth studies.

We simultaneously train neural ratio estimators $\nreapprox(\iparam, \data)$ for each individual parameter in a chosen model (e.g. $\iparam \in \qty{T_0, \log{g}, r_p, r_s, X_{\ch{H2O}}, X_{\ch{CO}}, X_{\ch{CO2}}, X_{\ch{H2S}}, X_{\ch{K}}, X_{\ch{Na}}, X_{\ch{SO2}}, \aerradd}$ in the TP1 CF model). After training, we evaluate each $\nreapprox(\iparam, \tdata)$ at the observed data and multiply by the respective prior density, $\prob(\iparam)$, to obtain a representation of the marginal posterior, $\prob(\iparam\given\tdata)$.

\subsubsection{Bayesian model selection with SBI}
While in traditional Bayesian pipelines model selection is typically understood as a separate task from parameter inference, in the context of SBI model selection can be seen as the pinnacle of marginal inference:  the target is the posterior probability distribution of the model itself, $\Model$, once all its parameters have been marginalized out: 
\begin{equation}\begin{split}\label{eqn:mpost}
    \prob(\Model\given\tdata)
    & \propto \prob(\Model) \times \prob(\tdata \given \Model) 
    \\ & \propto \prob(\Model) \times \int \prob(\tdata \given \nparams, \Model) \prob(\nparams \given \Model) \dd{\nparams}.
\end{split}\end{equation}
Here $\prob(\Model)$ is the prior model probability, and $\prob(\nparams \given \Model)$ and $\prob(\tdata \given \nparams, \Model)$ are, respectively, the prior and likelihood of \emph{all} the parameters of the model, which are once again implemented in a stochastic simulator. Notice that if the models have entirely different parameter spaces (e.g., different number of parameters in each, or differences in the prior distributions), this is effortlessly taken care of in the simulator.

In our approach, a neural network is trained to explicitly output the normalized posterior probabilities for all \nmod models considered. To this end, the training set combines simulations from each model in proportion to $\prob(\Model)$ (i.e., in equal numbers, since we adopt a uniform prior in the space of models). When a mock example \data is processed, the network's output that corresponds to the true model is maximized; but since \emph{the same} \data can be plausibly produced by \emph{multiple} models (due to noise or parameter degeneracies), the final result is proportional to the posterior probabilities for the models.
More formally, following \citet{Elsemuller_2023,Karchev-simsims}, we minimize the expected negative log-probability loss\footnote{This is similar to parameter inference with neural \emph{posterior} estimation, except that now the role of the parameter is taken by a discrete random variable that represents the model label. This simplifies the implementation of the necessary normalization, which can be performed explicitly over the discrete space.}
\begin{equation} \label{eq:model_sel_loss}
    - \sum_{m=1}^{\nmod} \prob(\Model=\aModel) \times \Expectation[\prob(\data\given\Model=\aModel)][\ln \opbraces{q_m}(\data)],
\end{equation}
where $q_m$ is the posterior probability that the NN assigns to $\aModel$, obtained by normalizing the raw network outputs $\qty[f_m]_{m=1}^{\nmod}$ via the softmax function:
\begin{equation}\label{eqn:softmax}
    q_m \equiv \frac{\exp(f_m)}{\sum_{m^\prime=1}^{\nmod} \exp(f_{m^\prime})}. 
\end{equation}

Notice that SBI delivers directly the posterior probabilities for all models, rather than the evidence $\evidence(\tdata) \equiv \prob(\tdata \given \Model=\aModel)$ for a single model, which is the result of a given likelihood-based analysis, e.g., with nested sampling. The set of evidences for all models can be converted to model probabilities according to equation (\ref{eqn:mpost}):
\begin{equation}\label{eqn:evidence-to-mpost}
    \prob(\Model=\aModel \given \tdata) = \frac{\evidence(\tdata) \prob(\Model=\aModel)}{\sum_m \evidence(\tdata) \prob(\Model=\aModel)}.
\end{equation}
Since this is of the same form as \cref{eqn:softmax}, the network outputs $f_m$ can be directly related to log-evidences (after training and \emph{when the prior over models is constant}):
\begin{equation}
    \opbraces{f_m}(\tdata) = \ln \evidence(\tdata) + \mathrm{constant}.
\end{equation}
Therefore, they can also be used to compute the \emph{Bayes factor} between any pair of models \aModel[i], \aModel[j]:
\begin{equation}\label{eqn:BF}
    \BF[ij](\tdata) \equiv \frac{\evidence[i](\tdata)}{\evidence[j](\tdata)} = \exp(\opbraces{f_i}(\tdata) - \opbraces{f_j}(\tdata)).
\end{equation}
The Bayes factor balances each model's ability to fit the data (likelihood) while accounting for their model complexity (prior volume), thus including a quantitative measure of Occam's razor \citep{trotta08}. Here, we follow Jeffreys' scale for the strength of evidence as presented in \citet{trotta08}:
\begin{equation*}
    \ln \BF[ij](\tdata) \begin{cases}
        \in \interval[open right]{1}{2.5} & \text{weak}, \\
        \in \interval[open right]{2.5}{5} & \text{moderate}, \\
        > 5 & \text{strong},
    \end{cases}
\end{equation*}
in favor of model \aModel[i].

\subsubsection{Advantages of amortized inference}

A key advantage of SBI over traditional Bayesian techniques like nested sampling is the so-called \emph{amortized} nature of its learning: this means that almost all of the computational effort is expended \emph{only once} to train the neural network for the tasks of interest (in our case, both parameter inference and model comparison), which are then carried out by querying the trained network -- which requires very little computation. In this sense, the initial computational cost is recouped (i.e., amortized) when re-using the trained network multiple times later (either on real or simulated data). 

In our case, the \acronym\ network is trained to perform inference on a large variety of possible data \data rather than for only a particular observed \tdata. Thus, after the upfront cost of training the network (broadly comparable to a single standard nested sampling run), inferences from \tdata (i.e. both marginal parameter posteriors and the posterior probabilities of all models) can be obtained at almost no computational cost via a single forward pass, which takes on the order of milliseconds. In contrast, even GPU-accelerated nested-sampling retrievals take \SI{\sim 10}{\hour} \emph{per model}. Importantly, SBI can be performed with similar speed on \emph{any} other data: e.g., further observations or mock examples, provided they are assumed to be plausible realizations from the simulator used for training, while nested sampling needs to be re-run from scratch every time.

Apart from the massive speed-up in inference that this entails, amortization enables several tests and calibrations that are normally impossible with traditional Bayesian methods. Analyses of a \emph{validation set} of simulated examples (which are not used for training) can be used to examine the Bayesian (i.e., averaged over parameters sampled from the priors) coverage properties of the approximated posteriors \citep[see also \citealp{Cook_2006} and \citealt{Talts_2020}]{Hermans_2022}. Another possibility is to construct from a large number of simulated reconstructions confidence regions with \emph{exact} and guaranteed frequentist coverage \citep[see also \citealp{Dalmasso_2020,Dalmasso_2022,Masserano_2022}]{Karchev-sicret}. In the case of model selection, one can test using a large number of simulated examples the \emph{reliability} of the posterior probabilities, as well as the ability of the network -- averaged over repeated draws from the parameters' priors -- to identify the correct model \citep[see also \citealp{Karchev-simsims}]{DeGroot_1983}. Finally, analyses of numerous simulated data allow investigation of the dependence of model-selection results on the true parameter values, which is computationally unfeasible in a traditional Bayesian setting as it would require a prohibitive number of nested-sampling retrievals. 

\subsubsection{Simulations, networks, and training}

For each of the $\nmod = 12$ models we consider, we simulate \num{e5} mock \emph{noiseless} spectra with parameters randomly sampled from the priors in \cref{tab:NS_priors}. Observational noise, according to equations (\ref{eqn:like}) and (\ref{eqn:noise}), is then added on the fly, i.e., a different realization is used in each training epoch, as a form of simple and fast data augmentation. Each training set separately is used to train a parameter-estimation network for the given model it has been generated from, while their concatenation (amounting to \num{1.2e6} examples) is used to train the model-selection network.

\begin{figure*}
\begin{subfigure}[b]{0.5\textwidth}
\centering
    \includegraphics{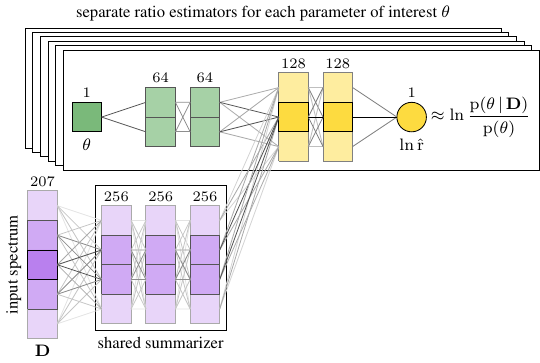}
    \caption{parameter-inference (NRE) network}
\end{subfigure}
\begin{subfigure}[b]{0.5\textwidth}
\centering
    \includegraphics{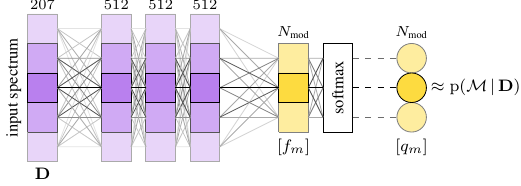}
    \caption{model-selection network}
\end{subfigure}
    \caption{Schematic of the \acronym\ inference networks, which we implement as multi-layer perceptions (MLPs). Panel (a): for a given choice of model, we train simultaneously ratio estimators $\nreapprox$ to approximate equation (\ref{eqn:nreratio}) for each parameter \iparam within that model. This allows evaluation of the marginal posteriors when queried with the observed data and multiplied by the respective priors $\prob(\iparam)$. Panel (b): a model-selection network is trained to estimate directly the (normalized) posterior probabilities of all $\nmod=12$ models considered in this paper.\label{fig:nn}}
\end{figure*}

We use simple fully connected neural networks (depicted in \cref{fig:nn}) for both parameter inference (NRE) and model selection. In both cases, the raw data \data is first pre-processed by a multi-layer perception (MLP) and embedded in \num{256} or \num{512} dimensions (for NRE and model selection, respectively). The resulting featurization is shared among the ratio estimators for all parameters (again implemented as MLPs), whereas for model selection it is directly converted into model probabilities by a linear layer of output size $\nmod=\num{12}$.

Generating \num{1.2e6} spectra took \SI{6}{\hour} on an NVIDIA RTX4090 GPU. The subsequent training\footnote{We train for 100 (fixed) epochs and use the checkpoint with the best validation loss.} took additionally about \SI{3}{\hour} per network. We emphasize that this upfront computational effort needs to be expended only once, after which the trained networks can be deployed on any data -- simulated or real, provided they are assumed to be a plausible sample from the simulator -- at almost no computational cost. 

\section{Results}
\label{sect:results}

\subsection{Validation on mock data}\label{sect:mock}

\begin{figure*}[!ht]
\begin{center}
\vspace{-0.1in} 
\includegraphics[width=2\columnwidth]{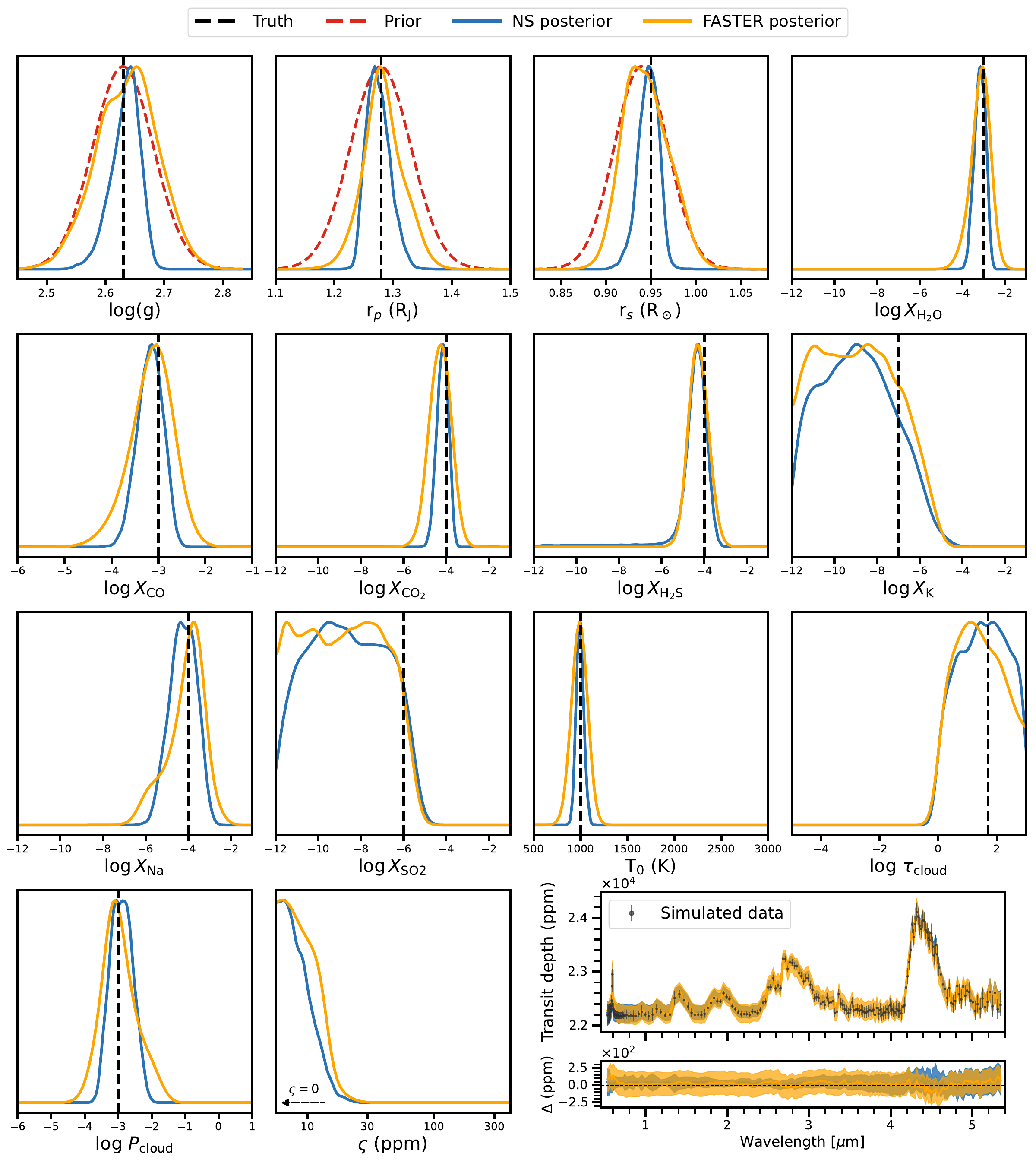}
\end{center}
\vspace{-0.1in}
\caption{Retrievals assuming the TP1 G model (isothermal profile and gray clouds) performed on a simulated spectrum from the same model and with parameter values indicated by black vertical dashed lines. The results (1-dimensional marginal posteriors, normalized to their peak) from nested sampling and SBI (blue and orange) are plotted as solid lines, while priors are depicted as dashed red lines, where not uniform across the plotted range. The bottom right panel shows the simulated data (error bars, containing only the reported instrumental noise, i.e., $\aerradd=0$ as simulated) and the reconstructions from the two methods, derived by simulating spectra from the posterior samples.}
\label{fig:posteriors_mock_g}
\end{figure*}

First of all, we wish to validate \acronym\ against the ground-truth results from \texttt{MultiNest} in a fully controlled environment, i.e., by analyzing a mock example. For this illustration, we assume the TP1 G model (isothermal with gray clouds) and simulate a noisy spectrum with parameter values as listed in \cref{tab:NS_priors}. No additional intrinsic variance is added to the measurement errors reported by JWST for WASP-39b, which corresponds to $\aerradd = 0$.

Figure \ref{fig:posteriors_mock_g} demonstrates that \acronym\ produces posterior distributions that are consistent with all of the input parameter values.  These posteriors are not only consistent with those generated by our nested sampling retrieval, but for several parameters the shapes of these posteriors are identical. This is remarkable particularly for parameters such as the chemical abundances and $\log \tau_\text{cloud}$ that can vary over several orders of magnitude. We remind the reader that our training is fully amortized across the whole prior range -- meaning that an almost instantaneous inference of the same quality can be obtained from the trained network for \emph{any} value of the underlying parameters across that range.

In the bottom right panel of \cref{fig:posteriors_mock_g}, we also show the reconstructed spectra from the two methods, compared to the simulated data. For this purpose, we sample \num{1000} parameter values according to the respective NS and SBI posteriors and plot the distribution (median and $\pm 1$ standard deviations) of the corresponding synthetic spectra. While NS gives access to the joint 13-dimensional posterior to sample from, with SBI we approximate only the 1-dimensional projections. To plot the SBI reconstruction, therefore, we form an approximation to the joint by multiplying the marginals; this necessarily leads to a wider (conservative) result, which is reflected in the greater spread in the reconstructions (see, in particular, at low wavelengths).

Next, we compare the results of likelihood- and simulation-based model selection, again on mock data: in addition to the example discussed above, we simulate a realization from the simpler cloudfree model (TP1 CF) with the same parameter values (where relevant). While the inference network directly delivers all $\nmod=\num{12}$ posterior probabilities simultaneously, with nested sampling we need to run 12 separate retrievals, each with a different model (and then convert the evidences to posterior probabilities via \cref{eqn:evidence-to-mpost}, assuming equal prior probabilities). In practical terms, this translates to approximately \SI{100}{\hour} of NS retrievals, compared to \SI{24}{\milli\second} for the single network evaluation.

\begin{figure*}[!ht]
\begin{center}
\vspace{-0.1in} 
\includegraphics[width=\columnwidth]{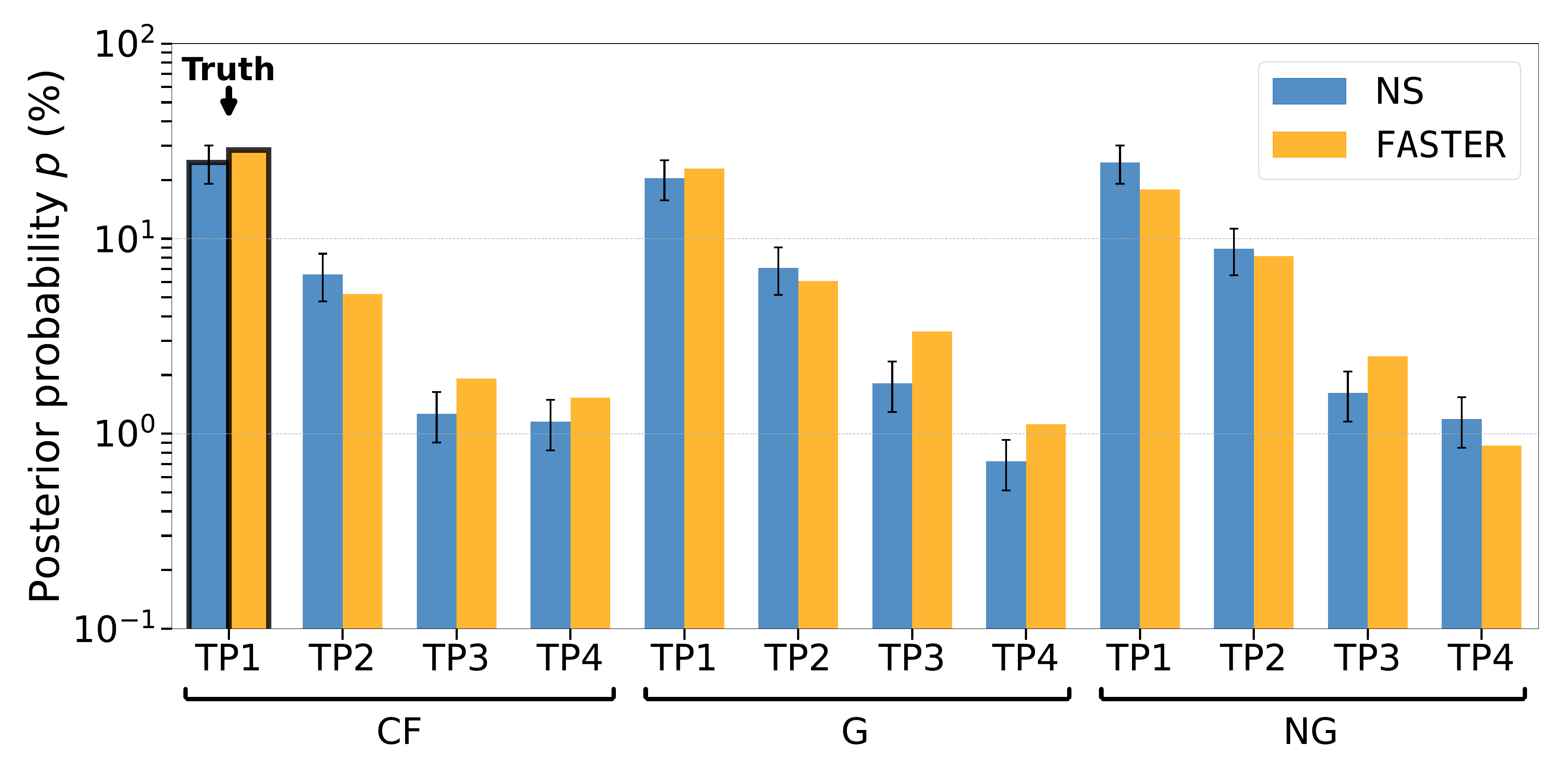}
\includegraphics[width=\columnwidth]{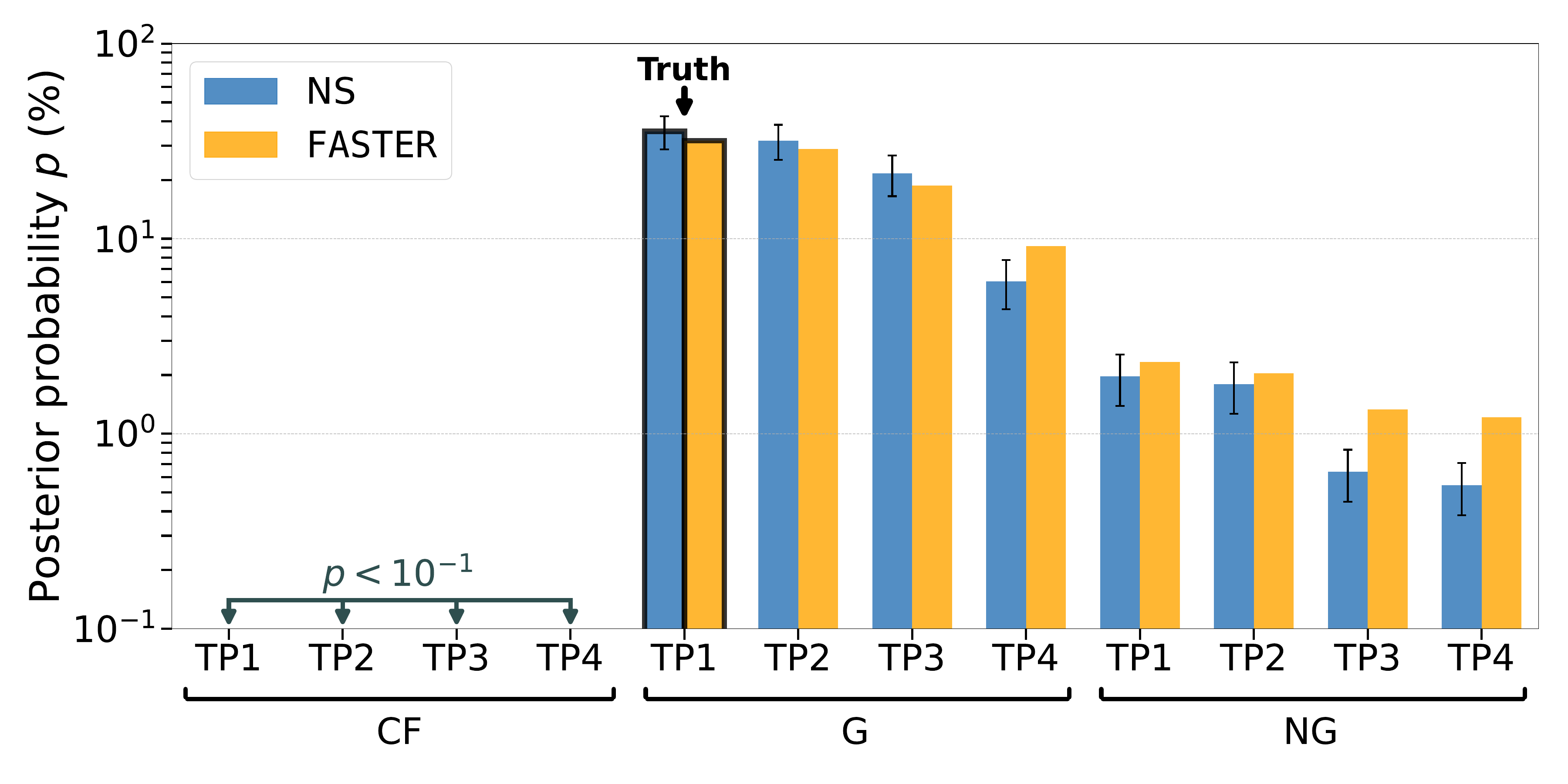}
\end{center}
\vspace{-0.2in}
\caption{Comparison of model posterior probabilities derived with SBI (orange) and nested sampling (blue) from mock data generated from the TP1 CF (isothermal, cloudfree: left panel) or TP1 G (isothermal, gray clouds: right panel) models.}
\label{fig:probabilities}
\end{figure*}

The results are compared in \cref{fig:probabilities}: we notice excellent agreement between the two approaches, with SBI probabilities falling generally within the estimated uncertainty from the NS runs. While there is stochasticity in the initialization of the neural network, once converged this is likely a subdominant source of error in the probability estimate from \acronym. A more important role is played by the ``systematics'' uncertainty coming from choice of network architecture, training epochs, batch size, learning rate, number of training examples, etc, which is difficult to estimate reliably. We therefore limit our output to a point estimate of the model probability. Additionally, when performing inference on real data using the likelihood-based nested sampling algorithm or NRE we can expect that any model mis-specification (i.e., when the real data are generated in a manner that is not exactly captured by our forward model) will lead to the resulting probability estimates to be different between the two methods. Concretely, in the presence of clouds (right panel: mock data from the TP1 G model), cloudfree models are decisively excluded, and moreover the type of clouds (gray/non-gray) is correctly identified. On the other hand, when the true model \emph{is} cloudfree (left panel), cloudy models cannot be excluded because their prior does include low values of $\tau_{\rm cloud}$, in which case the spectrum is indistinguishable from cloudfree. Moreover, whereas the temperature--pressure profile of a cloudfree atmosphere can be reliably identified, this is not the case when clouds are present. These points highlight the importance of priors on model-selection results and the latter's dependence on the underlying parameter values. We now proceed to explore these two effects further, employing the unique amortization property of SBI.

\subsection{Occam's razor and distinguishing cloud models\label{sec:results-occam}}

\begin{figure}
\begin{center}
\includegraphics[width=\columnwidth]{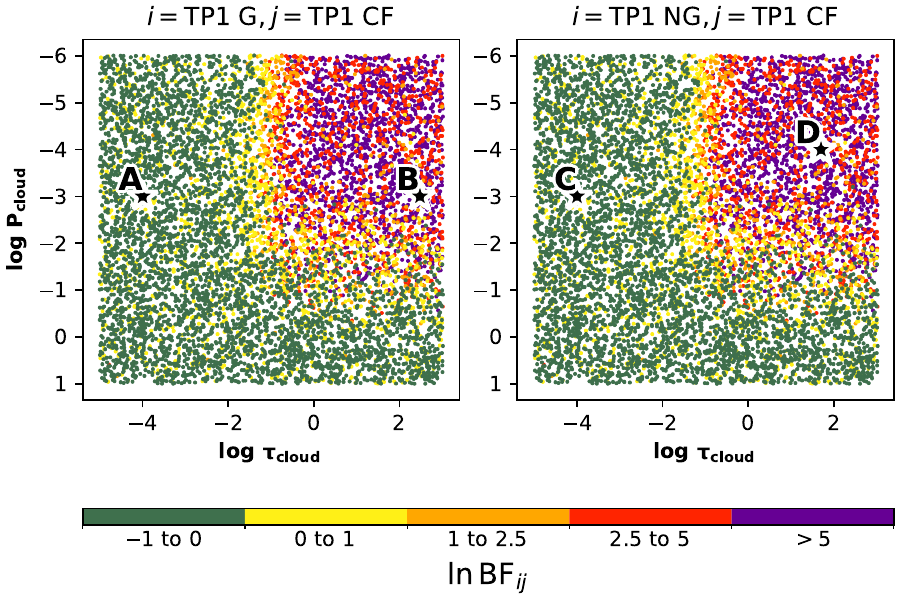}
\includegraphics[width=\columnwidth]{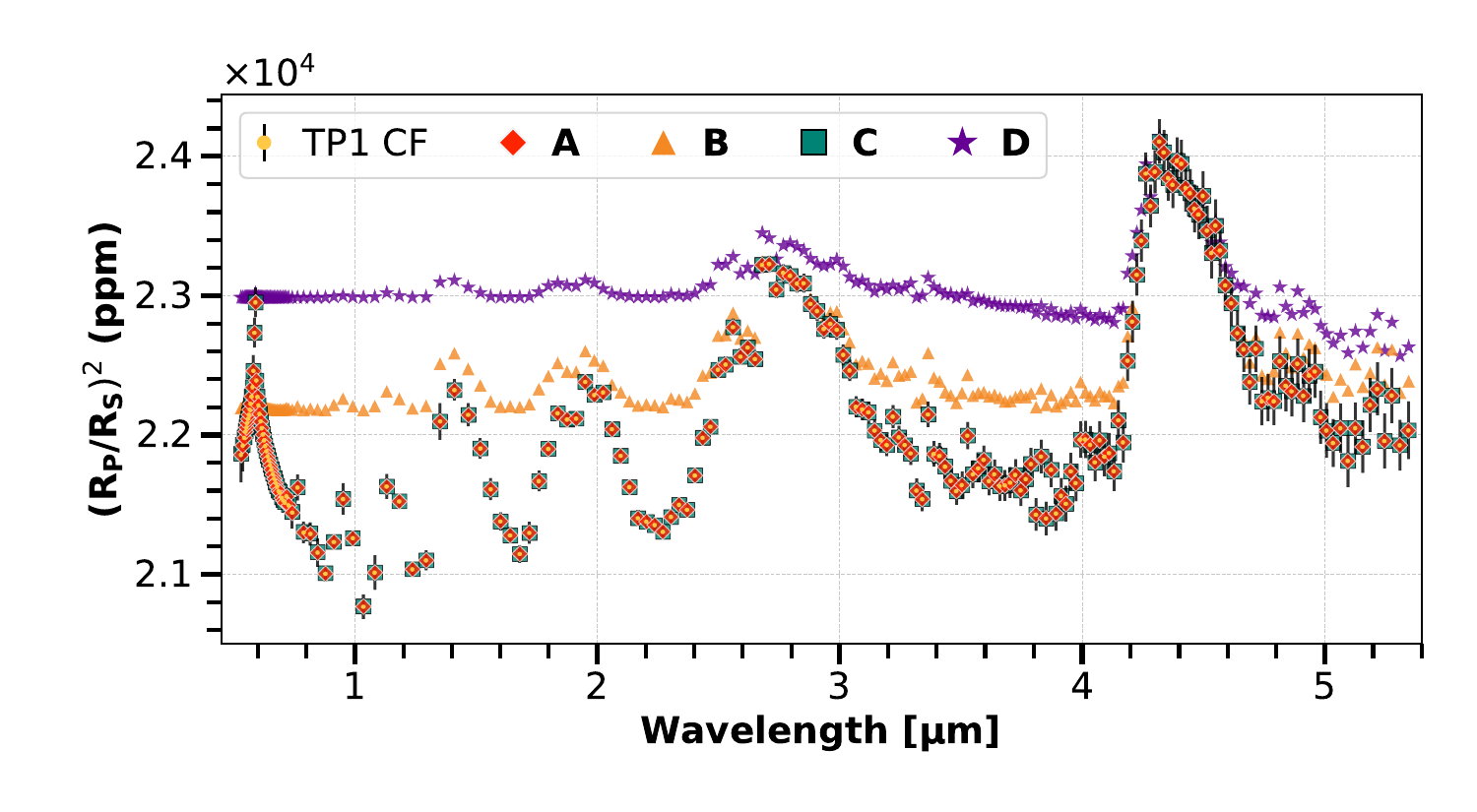}
\end{center}
\caption{Top panels: Bayes factor (color scale) comparing an isothermal cloudy (left: gray, right: non-gray) versus cloudfree model as a function of the cloud-top pressure ($P_{\rm cloud}$) and cloud optical depth ($\tau_{\rm cloud}$), for \num{9000} simulated data realizations. Parameters not shown are randomly sampled from their prior distributions for each realization. Bottom panel: synthetic spectra corresponding to the 4 realizations labeled in the top panels (and remaining parameters as in \cref{tab:NS_priors}). Also shown is a cloudfree spectrum with measurement errors reported by JWST for WASP-39b as illustration of the measurement uncertainty.}
\label{fig:clouds}
\end{figure}

Having validated SBI on concrete examples, we turn to investigating the results of Bayesian model selection as a general methodology (i.e., irrespective of the technique used). Still, the following demonstrations are only made possible by the amortization inherent in SBI, which allows results to be rapidly derived from numerous simulated examples.

Our first setup aims to determine the values of cloud-related parameters ($\tau_{\rm cloud}$ and $P_{\rm cloud}$) which -- for the given JWST noise levels -- \emph{would} lead to a positive identification of a cloudy versus cloudfree model. To this end, we simulate validation sets of \num{9000} mock spectra each from the gray- and non-gray cloud models\footnote{For this illustration, we focus on isothermal (TP1) models . The remaining parameters (including \aerradd) are again sampled from the priors in \cref{tab:NS_priors}.} and derive Bayes factors, via equation (\ref{eqn:BF}), from all of them with respect to the cloudfree model. While with nested sampling this extreme exercise would have cost \num{\sim e5} GPU-hours, it took us a mere \SI{0.6}{\second} in parallel on a single GPU.

The outcomes are plotted in the top panels of \cref{fig:clouds}, where the axes correspond to the \emph{true} parameters from which the mock data were generated, and the color scale reflects the NN-derived Bayes factors (equivalent to posterior odds since we take equal prior $\prob(\Model)$). The revealed pattern aligns well with our physical intuition about transmission spectra: when the cloud optical depth exceeds unity, Bayesian model comparison strongly ($\ln \BF[] >5$) favors a cloudy model.  (For the non-gray cloud model, recall that the optical depth is wavelength-dependent and referenced to \SI{1}{\micro\meter}, which is within the observed range.)  But this also depends on the cloud not residing too deeply in the atmosphere: if the cloud-top pressure resides at pressures greater than \SI{0.1}{\bar}, the cloud has a vanishingly small effect on the spectrum, and therefore, the cloudfree model is preferred by Bayesian model comparison in virtue of Occam's razor (yellow and green regions in the top panels).

In the bottom panel of \cref{fig:clouds}, visual inspection of particular spectra (labeled "A" and "C") confirms that the cloudy models are indistinguishable from one another and from the cloudfree case in the optically thin limit. The latter -- simpler -- model is both a sufficient and physically correct description of the data.  This failure to reject the cloudfree model, when it is indeed false, would be a ``type II'' error in the frequentist sense. However, from a Bayesian model comparison perspective this is indeed the correct conclusion: if the data do not allow to distinguish between two models (cloudfree vs. cloudy), selecting the simpler one is in accord with Occam's razor principle. We investigate this point on average across parameter values in the next section.

\subsection{Average performance of Bayesian model comparison\label{sec:results-refinedness}}

In the previous section, we scrutinized Bayesian model comparison for individual examples as a function of the underlying parameters. Here, we instead turn to the results \emph{averaged} across parameters sampled from each model's prior distributions and examine the "expected model posterior probability" (EMPP) matrix.\footnote{While this appears similar to the standard confusion matrix used in multi-class classification, it is a different construct in that it shows average posterior probabilities across its rows rather than classification frequencies. In \citet{Karchev-simsims}, following \citet{DeGroot_1983}, it is referred to as the "refinedness" matrix of the probabilistic classifier.} Its construction again relies on the amortized nature of SBI and is otherwise impossible using traditional nested sampling retrievals.

\begin{figure}[!ht]
\centering
\includegraphics[width=0.8\columnwidth]{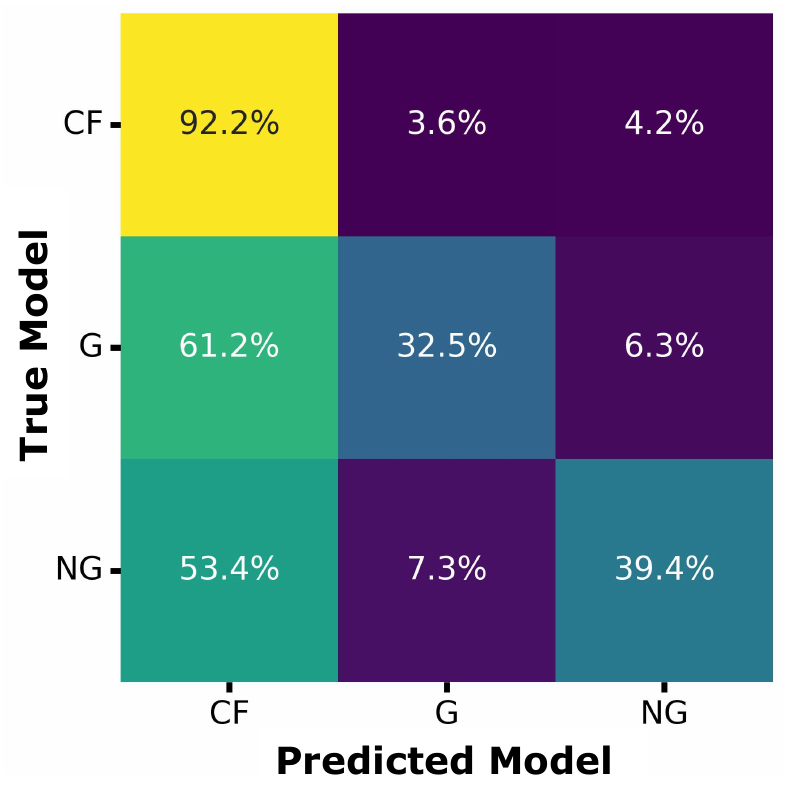}
\includegraphics[width=0.81\columnwidth]{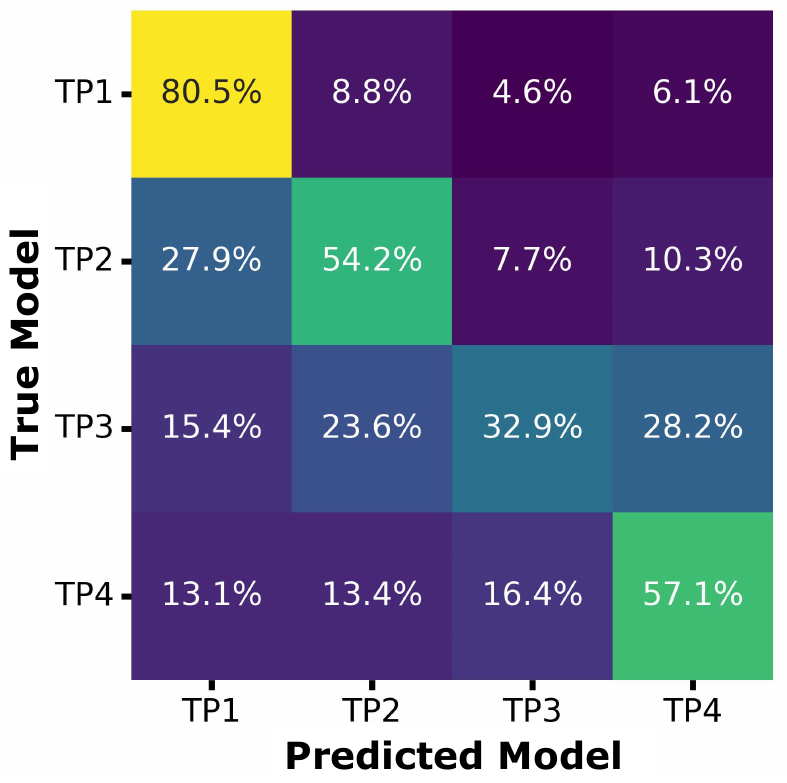}
\vspace{0.1in}
\caption{Expected model posterior probability (EMPP) matrix for cloud models (top panel; cloudfree, gray clouds, non-gray clouds from top to bottom) and temperature--pressure profile models (bottom panel, increasing the number of slope changes in the profile from top to bottom). Each row is the average posterior probability distribution (over the models as labeled on the bottom) when evaluated for random examples from the model given on the left.}
\label{fig:EMPP_matrix}
\end{figure}

In detail, we simulate \num{9000} mock realizations from a given model (with parameters drawn from the priors as above (including \aerradd) and evaluate the trained model-selection network on them to obtain the corresponding posterior probability distributions $\qty[q_m]$. Averaging them across the \num{9000} examples produces a single row of the EMPP matrix; repeating this with each of the 12 models gives the full \numproduct{12x12} matrix depicted in \cref{fig:LargeMatrix}. For clarity, we further group models according to temperature--pressure profile (TP1--4) or cloud type (CF, G, NG) and obtain the matrices shown in \cref{fig:EMPP_matrix}.

The interpretation of an EMPP matrix is as follows. The values on the diagonal represent the probabilities assigned to the "correct" model, i.e., the one that data were generated from, and so it is in general desirable that these entries are close to unity. However, since a given data set may have been plausibly produced by several models, either due to noise fluctuations or parameter degeneracies, non-zero off-diagonal entries are an expected feature of Bayesian model comparison and arise under two different circumstances.

Firstly, for nested models (e.g., $\mathrm{CF} \subset \mathrm{G} \subset \mathrm{NG}$), when the strength of the "additional" effect (e.g., the optical depth $\tau_{\rm cloud}$ of the cloud layer) is negligible in comparison with the noise, Bayesian model comparison includes the principle of Occam's razor and favors the simpler (nested) model, as demonstrated in \cref{fig:clouds}. Here, it manifests in the prominent entries below the diagonal\footnote{Notice that models in the EMPP matrix are arranged from top to bottom and left to right in order of increasing complexity.} of the top panel of \cref{fig:EMPP_matrix}, which compares all three cloud models on average, and similarly for the temperature--pressure profiles in the bottom panel.

Secondly, it is also possible that significant posterior probability is assigned to models that are more complicated than the true one: e.g., TP3\textrightarrow TP4 in the bottom panel of \cref{fig:EMPP_matrix}, when the data is not sufficiently informative to distinguish between them. Consider the extreme case in which the models cannot be distinguished at all: then, the posterior probabilities revert to the prior, which in our case is equal among all models. This can happen either because the data is \emph{uninformative} about the two models (e.g., the noise level is much larger than the differences in the spectra predicted by the two models) or because the network is not optimally trained. In general, it is difficult to distinguish the source of such probability leakage, but given that the SBI results we examined in \cref{fig:probabilities} are very close to the ground truth, we tend to exclude network's undertraining as the main culprit. Regardless, the EMPP matrix quantifies the model comparison average performance using SBI irrespective of the source of the probability leakage to the upper right off-diagonal elements.

Finally, we remark that the EMPP matrix is a useful tool to evaluate the effectiveness of model comparison. It results from a complex interplay between quality of the data (the observational noise); completeness and accuracy of the models being compared; residual intrinsic dispersion; and prior ranges for the models' parameters (which control the strength of the Occam's razor effect). To our knowledge, no other approach but amortized SBI can deliver this kind of insight with reasonable computational effort.  

\subsection{Application to JWST data of WASP-39b}

Finally, we test the \acronym\ framework on real JWST spectra.  Previously, \cite{lueber24} performed nested sampling retrievals on the NIRSpec PRISM spectrum (\SIrange[range-phrase=--, range-units=single]{0.53}{5.34}{\micro\meter}) of the hot Jupiter WASP-39b. They determined that the isothermal gray-cloud model (TP1 G in the terminology of the current study) has the highest Bayesian evidence. \acronym\ reproduces this conclusion with the following ranking of posterior probabilities: TP1 G (26.7\%), TP2 G (21.1\%), TP4 G (18.1\%), TP3 G (17.6\%); non-gray clouds and cloudfree models have posterior probabilities of at most 4.2\% and 2.4\%, respectively.

\begin{figure*}[!ht]
\begin{center}
\includegraphics[width=2\columnwidth]{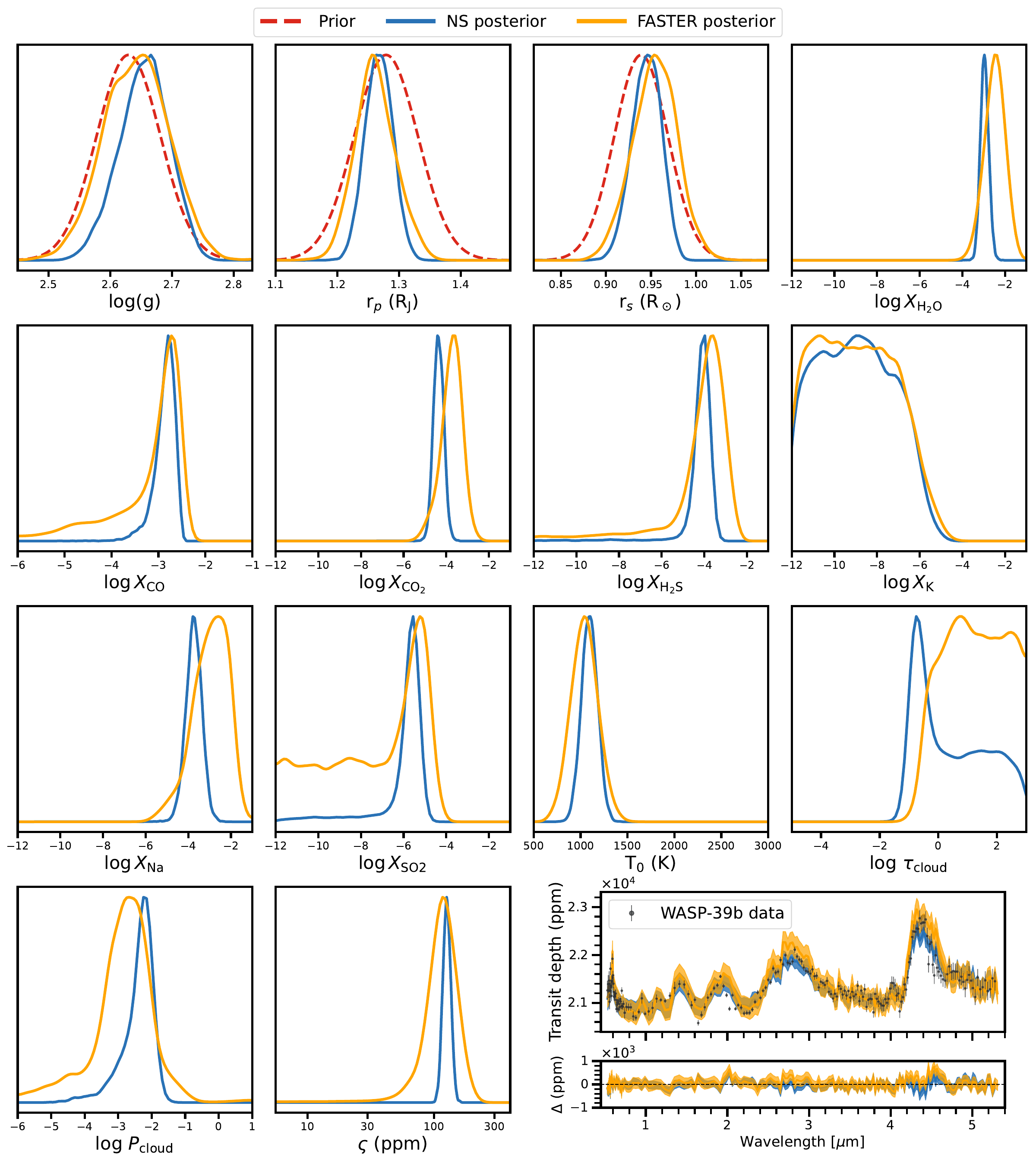}
\end{center}
\vspace{-0.1in}
\caption{Atmospheric retrieval analysis of the JWST NIRSpec PRISM spectrum of the hot Jupiter WASP-39b using both \acronym\ and nested sampling for the TP1 G model preferred by Bayesian model comparison.  The nested sampling analysis was previously published by \cite{lueber24} and reproduced here for convenience of comparison.  The marginal posterior distributions computed by SBI and nested sampling are shown in orange and blue, respectively, normalized to their peak. Priors (dashed red) are shown only where not uniform across the range. In the spectrum panel (bottom right), only the measurement uncertainty of the data is shown (without the additional intrinsic dispersion, which is however accounted for in the fit). The SBI and nested sampling retrievals required $\sim1$ s (post-training) and $\sim8$ hours of GPU time, respectively.}
\label{fig:posteriors_wasp39b}
\end{figure*}

Figure \ref{fig:posteriors_wasp39b} displays the marginal posterior distributions of parameters obtained using both nested sampling and SBI, restricting the analysis to the TP1 G model that is preferred by Bayesian model comparison. There is good agreement between the median values and shapes of the posteriors, but the differences here are more pronounced than in the case of simulated data (\cref{fig:posteriors_mock_g}). Curiously, the residuals between data and model are most pronounced around \SI{3}{\micro\meter} and \SI{4.4}{\micro\meter}, where the spectral features of water and carbon dioxide are expected to dominate the fluxes. 

Moreover, we find -- as \citet{lueber24} did with nested sampling -- evidence for additional scatter present in the data at the $\aerradd \approx \SI{100}{ppm}$ level. This implies that the physical and/or instrumental modeling (implemented in the simulator \emph{and} the nested-sampling likelihood) may be incomplete and therefore the real data may not be a plausible sample from the model, i.e., it may lie outside the distribution of training examples; therefore, we should not expect a perfect match between SBI and likelihood-based results in this case, as model misspecification has in general different failure modes in likelihood-based inference and SBI. Verifying that the observed data are in-distribution with respect to the training set is a key open issue that is the subject of ongoing research in the SBI community (see e.g., \citealt{Wehenkel2024}). 

Despite the above, this exercise demonstrates that our \acronym\ approach produces essentially equivalent scientific conclusions to what can be obtained via nested sampling.

\section{Discussion}
\label{sect:discussion}

\subsection{Comparison to previous work} 

A few recent studies in the exoplanet literature use alternative SBI approaches to atmospheric retrieval. \Citet{vasist23} apply neural posterior estimation (NPE) for computationally efficient retrievals, using normalizing flows to estimate the posterior in an amortized way. NPE is also used by \cite{aubin23}, the winner of the ARIEL 2023 data challenge\footnote{\url{https://www.ariel-datachallenge.space/adc2023/}} and \citet{gebhard25}, who compare discrete and continuous normalizing flows and refine their results with (likelihood-based) importance sampling for improved posteriors and Bayesian evidence estimates. Lastly, \citet{ardevolmartinez24} estimate posteriors sequentially, re-using initial approximate results to simulate training data targeted to each observation. While this improves precision, it sacrifices amortization.

In contrast to these studies, we have employed neural ratio estimation for parameter inference, which relies on a very simple feed-forward neural network instead of a normalizing flow. While we have only estimated one-dimensional marginals---which are usually sufficient for drawing scientific conclusions---ratio estimation can be extended to derive the joint posterior over many ($\gtrsim 20$) parameters with essentially the same simulation and training-time requirements \citep{AnauMontel_2024}. Furthermore, the likelihood-to-evidence ratio is independent of the prior (modulo parameter-independent normalization), which allows posterior evaluation under different prior assumptions and fully amortized sequential training, which we will demonstrate in future work. Moreover, we have presented the first simulation-based fully Bayesian and amortized model comparison of models of exoplanetary atmospheres, which bypasses explicit evidence calculations and directly compares (and ranks) numerous alternatives. This technique, while distinct from the simulation-based approaches to parameter estimation discussed above, can fully reuse existing simulations and still relies on a simple classifier network architecture. Finally, we have applied both methods to the real JWST PRISM spectrum of WASP-39b and obtained results in good agreement with traditional nested-sampling retrievals.

\subsection{Implications and opportunities for future work}

While this study validated the SBI approach for a specific exoplanet, with minimal additional training effort it is possible to generalize our networks to be applicable for inference and model selection to a broad range of planetary radii, stellar radii and planetary surface gravity. On top of one-dimensional marginal posteriors, the same inference strategy can be trivially extended to two-dimensional joint marginal posteriors for all pairs of parameters of interest. Exploiting the amortized nature of the network, posterior distributions can be calibrated to obtain exact frequentist intervals with guaranteed coverage, following the procedure described in \cite{Karchev-sicret}.

With such a tool in hand, it becomes possible to use it for near-instantaneous retrieval over a large ($\gtrsim 100$) sample of spectra, and to extend it to an ensemble ($\gtrsim 1000$) of models. For each exoplanet, one could trivially perform Bayesian model averaging, where the final retrieval becomes a weighted average of every model (according to its respective posterior probability) in the ensemble. Additionally, population-level studies become feasible without compromising on any aspect of the analysis -- including selection effects, which are key to obtaining accurate inferences for the population parameters of exoplanets. 

The computational streamlining and greatly increased efficiency of the SBI framework will also allow to finally incorporate the uncertainties associated with the cross sections or opacities of atoms and molecules into atmospheric retrievals.  Using traditional methods such as nested sampling, this improvement would be computationally prohibitive as each retrieval would incur order-of-magnitude increases in computational cost.  Using SBI, we would merely require that the empirical uncertainties associated with these cross sections be stochastically sampled in the simulator.

We believe that SBI approaches of the kind presented here will constitute as large a jump forward in spectral analysis of exoplanetary atmospheres as the original retrieval methods based on Markov Chain Monte Carlo have proven to be. 

\vspace{0.1in}

{\scriptsize RT acknowledges co-funding from Next Generation EU, in the context of the National Recovery and Resilience Plan, Investment PE1 – Project FAIR ``Future Artificial Intelligence Research''. This resource was co-financed by the Next Generation EU [DM 1555 del 11.10.22]. RT is partially supported by the Fondazione ICSC, Spoke 3 ``Astrophysics and Cosmos Observations'', Piano Nazionale di Ripresa e Resilienza Project ID CN00000013 ``Italian Research Center on High-Performance Computing, Big Data and Quantum Computing'' funded by MUR Missione 4 Componente 2 Investimento 1.4: Potenziamento strutture di ricerca e creazione di ``campioni nazionali di R\&S (M4C2-19 )'' - Next Generation EU (NGEU). 
AL acknowledges partial financial support from the Swiss National Science Foundation (via grant number 192022 awarded to KH). 
KH and MH acknowledge partial financial support from the  European Research Council (ERC) Geoastronomy Synergy Grant (grant number 101166936). CF acknowledges financial support from the ERC under the European Union’s Horizon 2020 research and innovation program (grant agreement number 805445).}

{\scriptsize AL and KK performed all numerical calculations, created the figures and led all aspects of the technical implementation.  CF, RT and KH initiated the model design of the project and provided scientific direction. MH assisted in the calculations, as part of his Master project training, and checked for reproducibility. All co-authors participated in a long series of Zoom discussions and contributed to the writing of the manuscript. No generative AI was used in the production of this manuscript.}

\appendix

\section{Full EMPP matrix}

\begin{figure*}[ht]
\begin{center}
\includegraphics[width=\columnwidth]{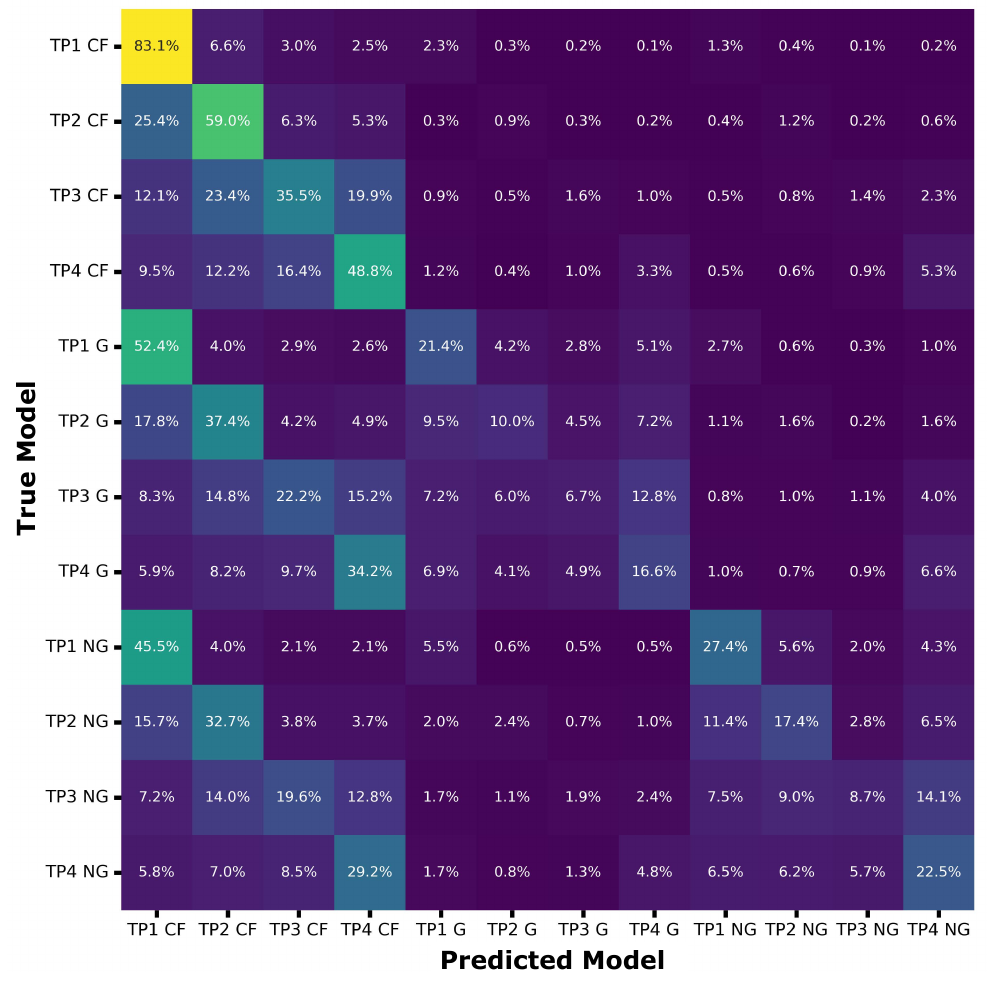}
\end{center}
 \vspace{-0.1in}
\caption{The full expected model posterior probability (EMPP) matrix. See \cref{sec:results-refinedness} for construction and interpretation.\label{fig:LargeMatrix}}
\vspace{-0.15in}
\label{fig:MS_diagnostics}
\end{figure*}

For completeness, we provide the full $12 \times 12$ EMPP matrix in Figure \ref{fig:MS_diagnostics}. From this matrix, the two marginalized matrices in Figure~\ref{fig:EMPP_matrix} can be derived by summing over the temperature profile model (thus obtaining the top matrix in Figure~\ref{fig:EMPP_matrix}) or over the cloud type model (thus obtaining the lower matrix). We notice in this matrix the appearance of diagonal substructure in the lower-diagonal block sub-matrices spanning the same cloud class (e.g., TP1 G to TP4 G sub-matrix): similarly to what was discussed in the main text, those are instances of Occam's razor preferring a simpler model (e.g., TP2 CF when the true model is TP2 G as the data are insufficiently constraining. 

\begin{thebibliography}{99}  

\bibitem[Abel et al.(2011)]{abel2011} Abel, M., Frommhold, L., Li, X., et al.\ 2011, Journal of Physical Chemistry A, 115, 6805. doi:10.1021/jp109441f

\bibitem[Abel et al.(2012)]{abel2012} Abel, M., Frommhold, L., Li, X., et al.\ 2012, \jcp, 136, 044319. doi:10.1063/1.3676405


\bibitem[Anau Montel et al.(2024)]{AnauMontel_2024} Anau Montel, N., Alvey, J., \& Weniger, C.\ 2024, \mnras, 530, 4107. doi:10.1093/mnras/stae1130


\bibitem[Ard{\'e}vol Mart{\'\i}nez et al.(2022)]{ardevolmartinez22} Ard{\'e}vol Mart{\'\i}nez, F., Min, M., Kamp, I., \& Palmer, P. I.\ 2022, \aap, 662, A108. doi:10.1051/0004-6361/202142976

\bibitem[Ard{\'e}vol Mart{\'\i}nez et al.(2024)]{ardevolmartinez24} Ard{\'e}vol Mart{\'\i}nez, F., Min, M., Huppenkothen, D., et al.\ 2024, \aap, 681, L14. doi:10.1051/0004-6361/202348367

\bibitem[Aubin et al.(2023)]{aubin23} Aubin, M., Cuesta-Lazaro, C., Tregidga, E., et al.\ 2023, arXiv:2309.09337. doi:10.48550/arXiv.2309.09337

\bibitem[Azzam et al.(2016)]{azzam2016} Azzam, A.~A.~A., Tennyson, J., Yurchenko, S.~N., et al.\ 2016, \mnras, 460, 4063. doi:10.1093/mnras/stw1133

\bibitem[Barstow \& Heng(2020)]{bh20} Barstow, J.~K. \& Heng, K.\ 2020, \ssr, 216, 82. doi:10.1007/s11214-020-00666-x

\bibitem[Benneke \& Seager(2012)]{bs12} Benneke, B. \& Seager, S.\ 2012, \apj, 753, 100. doi:10.1088/0004-637X/753/2/100

\bibitem[Benneke \& Seager(2013)]{bs13} Benneke, B. \& Seager, S.\ 2013, \apj, 778, 153. doi:10.1088/0004-637X/778/2/153

\bibitem[Brown(2001)]{brown01} Brown, T.M. \ 2001, \apj, 553, 1006

\bibitem[Carter et al.(2024)]{carter24} Carter, A.L., May, E.M., Espinoza, N., et al. \ 2024, Nature Astronomy, 8, 1008

\bibitem[Cobb et al.(2019)]{cobb19} Cobb, A.~D., Himes, M.~D., Soboczenski, F., et al.\ 2019, AJ, 158, 33.
doi:10.3847/1538-3881/ab2390


\bibitem[Cole et al.(2022)]{Cole_2022} Cole, A., Miller, B.~K., Witte, S.~J., et al.\ 2022, \jcap, 2022, 004. doi:10.1088/1475-7516/2022/09/004


\bibitem[Cook et al.(2006)]{Cook_2006} Cook, S.~R., Gelman, A., \& Rubin, D.~B.\ 2006, Journal of Computational and Graphical Statistics, 15, 675. doi:10.1198/106186006X136976

\bibitem[Cranmer et al.(2020)]{Cranmer_2020} Cranmer, K., Brehmer, J., \& Louppe, G.\ 2020, Proceedings of the National Academy of Science, 117, 30055. doi:10.1073/pnas.1912789117

\bibitem[Dalmasso et al.(2020)]{Dalmasso_2020} Dalmasso, N., Izbicki, R., \& Lee, A.~B.\ 2020, arXiv:2002.10399. doi:10.48550/arXiv.2002.10399

\bibitem[Dalmasso et al.(2022)]{Dalmasso_2022} Dalmasso, N., Masserano, L., Zhao, D., et al.\ 2022, arXiv:2107.03920. doi:10.48550/arXiv.2107.03920

\bibitem[DeGroot \& Fienberg(1983)]{DeGroot_1983} DeGroot, M.~H., \& Fienberg, S.~E.\ 1983, Journal of the Royal Statistical Society. Series D (The Statistician), 32, 12. doi:10.2307/2987588


\bibitem[Elsem{\"u}ller et al.(2023)]{Elsemuller_2023} Elsem{\"u}ller, L., Schnuerch, M., B{\"u}rkner, P.-C., et al.\ 2023, arXiv:2301.11873. doi:10.48550/arXiv.2301.11873

\bibitem[Feroz et al.(2009)]{feroz09} Feroz, F., Hobson, M.~P., \& Bridges, M.\ 2009, \mnras, 398, 1601. doi:10.1111/j.1365-2966.2009.14548.x

\bibitem[Fortney(2005)]{fortney05} Fortney, J.J. \ 2005, MNRAS, 364, 649

\bibitem[Fu et al.(2025)]{fu25} Fu, G., Stevenson, K.B., Sing, D.K., et al. \ 2025, ApJ, in press (arXiv:2501.02081)

\bibitem[Gebhard et al.(2025)]{gebhard25} Gebhard, T.~D., Wildberger, J., Dax, M., et al.\ 2025, \aap, 693, A42. doi:10.1051/0004-6361/202451861

\bibitem[Gelman(2014)]{Gelman2014} 
Andrew Gelman, John B. Carlin, Hal S. Stern, David B. Dunson, Aki Vehtari, and Donald B. Rubin,  \textit{Bayesian Data Analysis}, Third edition, CRC Press, 2014.

\bibitem[Grimm \& Heng(2015)]{gh15} Grimm, S.~L. \& Heng, K.\ 2015, \apj, 808, 182. doi:10.1088/0004-637X/808/2/182

\bibitem[Grimm et al.(2021)]{grimm21} Grimm, S.~L., Malik, M., Kitzmann, D., et al.\ 2021, \apjs, 253, 30. doi:10.3847/1538-4365/abd773

\bibitem[Hermans et al.(2019)]{Hermans_2020} Hermans, J., Begy, V., \& Louppe, G.\ 2019, arXiv:1903.04057. doi:10.48550/arXiv.1903.04057

\bibitem[Hermans et al.(2022)]{Hermans_2022} Hermans, J., Delaunoy, A., Rozet, F., et al.\ 2022, Transactions on Machine Learning Research, arXiv:2110.06581. doi:10.48550/arXiv.2110.06581

\bibitem[Karchev et al.(2023)]{Karchev-sicret} Karchev, K., Trotta, R., \& Weniger, C.\ 2023, \mnras, 520, 1056. doi:10.1093/mnras/stac3785

\bibitem[Karchev et al.(2023)]{Karchev-simsims} Karchev, K., Trotta, R., \& Weniger, C.\ 2023, arXiv:2311.15650. doi:10.48550/arXiv.2311.15650

\bibitem[Kirk et al.(2025)]{kirk25} Kirk, J., Ahrer, E.-M., Claringbold, A.B., et al. \ 2025, MNRAS, 537, 3027

\bibitem[Kitzmann \& Heng(2018)]{kh18} Kitzmann, D. \& Heng, K.\ 2018, \mnras, 475, 94. doi:10.1093/mnras/stx3141

\bibitem[Kitzmann et al.(2020)]{kitzmann20} Kitzmann, D., Heng, K., Oreshenko, M., et al.\ 2020, \apj, 890, 174. doi:10.3847/1538-4357/ab6d71

\bibitem[Lee et al.(2013)]{lee13} Lee, J.-M., Heng, K., \& Irwin, P.~G.~J.\ 2013, \apj, 778, 97. doi:10.1088/0004-637X/778/2/97

\bibitem[Li et al.(2015)]{li2015} Li, G., Gordon, I.~E., Rothman, L.~S., et al.\ 2015, \apjs, 216, 15. doi:10.1088/0067-0049/216/1/15

\bibitem[Line et al.(2013)]{line13} Line, M.~R., Wolf, A.~S., Zhang, X., et al.\ 2013, \apj, 775, 137. doi:10.1088/0004-637X/775/2/137

\bibitem[Line et al.(2015)]{line15} Line, M.~R., Teske, J., Burningham, B., et al.\ 2015, \apj, 807, 183. doi:10.1088/0004-637X/807/2/183

\bibitem[Lueber et al.(2024)]{lueber24} Lueber, A., Novais, A., Fisher, C., et al.\ 2024, \aap, 687, A110. doi:10.1051/0004-6361/202348802

\bibitem[Lueckmann et al.(2018)]{Lueckmann_2019} Lueckmann, J.-M., Bassetto, G., Karaletsos, T., et al.\ 2018, arXiv:1805.09294. doi:10.48550/arXiv.1805.09294

\bibitem[Lueckmann et al.(2021)]{Lueckmann_2021} Lueckmann, J.-M., Boelts, J., Greenberg, D.~S., et al.\ 2021, arXiv:2101.04653. doi:10.48550/arXiv.2101.04653

\bibitem[MacKay(2003)]{MacKay2003} 
David J. C. MacKay, \textit{Information Theory, Inference, and Learning Algorithms}, Cambridge University Press, 2003.  

\bibitem[Madhusudhan \& Seager(2009)]{ms09} Madhusudhan, N. \& Seager, S.\ 2009, \apj, 707, 24. doi:10.1088/0004-637X/707/1/24

\bibitem[Mancini et al.(2018)]{mancini18} Mancini, L., Esposito, M., Covino, E., et al.\ 2018, \aap, 613, A41. doi:10.1051/0004-6361/201732234

\bibitem[M{\'a}rquez-Neila et al.(2018)]{mn18} M{\'a}rquez-Neila, P., Fisher, C., Sznitman, R., et al.\ 2018, Nature Astronomy, 2, 719. doi:10.1038/s41550-018-0504-2

\bibitem[Masserano et al.(2022)]{Masserano_2022} Masserano, L., Dorigo, T., Izbicki, R., et al.\ 2022, arXiv:2205.15680. doi:10.48550/arXiv.2205.15680

\bibitem[Papamakarios \& Murray(2016)]{Papamakarios_2016} Papamakarios, G. \& Murray, I.\ 2016, arXiv:1605.06376. doi:10.48550/arXiv.1605.06376

\bibitem[Papamakarios et al.(2018)]{Papamakarios_2019} Papamakarios, G., Sterratt, D.~C., \& Murray, I.\ 2018, arXiv:1805.07226. doi:10.48550/arXiv.1805.07226

\bibitem[Polyansky et al.(2018)]{polyansky2018} Polyansky, O.~L., Kyuberis, A.~A., Zobov, N.~F., et al.\ 2018, \mnras, 480, 2597. doi:10.1093/mnras/sty1877

\bibitem[Skilling(2006)]{skilling06} Skilling, J. \ 2006, Bayesian Analysis, 1, 833

\bibitem[Talts et al.(2018)]{Talts_2020} Talts, S., Betancourt, M., Simpson, D., et al.\ 2018, arXiv:1804.06788. doi:10.48550/arXiv.1804.06788

\bibitem[Tashkun \& Perevalov(2011)]{tashkun2011} Tashkun, S.~A. \& Perevalov, V.~I.\ 2011, \jqsrt, 112, 1403. doi:10.1016/j.jqsrt.2011.03.005

\bibitem[Trotta(2008)]{trotta08} Trotta, R.\ 2008, Contemporary Physics, 49, 71. doi:10.1080/00107510802066753

\bibitem[Underwood et al.(2016)]{underwood2016} Underwood, D.~S., Tennyson, J., Yurchenko, S.~N., et al.\ 2016, \mnras, 459, 3890. doi:10.1093/mnras/stw849

\bibitem[Vasist et al.(2023)]{vasist23} Vasist, M., Rozet, F., Absil, O., et al.\ 2023, \aap, 672, A147. doi:10.1051/0004-6361/202245263

\bibitem[Waldmann et al.(2015)]{waldmann15} Waldmann, I.~P., Tinetti, G., Rocchetto, M., et al.\ 2015, \apj, 802, 107. doi:10.1088/0004-637X/802/2/107

\bibitem[Wehenkel et al. (2024)]{Wehenkel2024}
Wehenkel, A., Gamella, J. L., Sener, O., Behrmann, J., Sapiro, G., Cuturi, M., \& Jacobsen, J.-H.  
{\em arXiv}, abs/2405.08719, 2024.  

\bibitem[Yip et al.(2021)]{yip21} Yip, K.~H., Changeat, Q.,  Nikolaou, N., et al.\ 2021, AJ, 162, 195. 
doi: 10.3847/1538-3881/ac1744

\bibitem[Yip et al.(2024)]{yip24} Yip, K.~H., Changeat, Q., Al-Refaie, A., et al.\ 2024, \apj, 961, 30. doi:10.3847/1538-4357/ad063f


\bibitem[Zingales \& Waldmann(2018)]{zw18} Zingales, T. \& Waldmann, I.~P.\ 2018, \aj, 156, 268. doi:10.3847/1538-3881/aae77c

\end{thebibliography}
\end{document}